\documentclass[12pt]{article}%
\usepackage{amsmath}
\usepackage{graphicx}%
\usepackage{amsfonts}%
\usepackage{amssymb}
\def\ba{\begin{eqnarray}}
\def\ea{\end{eqnarray}}
\def\ll{\label}
\def\lb{\label}
\def\nn{\nonumber \\}
\def\bi{\bibitem}
\def\d{\delta}

\def\s{\sigma}

\def\rr{\rightarrow}
\def\p{p_\perp}

\def\m{m_\perp}
\def\mm{\mu_\perp}
\begin{document}

\title{ Wounded constituents}

\author{A.Bialas  \\ H.Niewodniczanski Institute of Nuclear Physics\\
Polish Academy of Sciences\thanks{Address: Radzikowskiego 152, Krakow,
Poland}\\and\\
M.Smoluchowski Institute of Physics \\Jagellonian
University\thanks{Address: Reymonta 4, 30-059 Krakow, Poland;
e-mail:bialas@th.if.uj.edu.pl;}}

\maketitle

keywords: wounded nucleons, hadronic constituents, nuclear enhancement

\begin{abstract}

The concept of the "wounded" hadronic constituents is formulated. Preliminary estimates indicate  that it may help to understand the transverse mass dependence of the particle production in hadron-nucleus and nucleus-nucleus collisions.

\end{abstract}

\section{Introduction}

The concept of a "wounded" source of particles, formulated long time ago
\cite{bbc,bcf}, turned out  useful in description of particle
production from nuclear targets at low transverse momentum \cite{busza}.
In this note, after recalling the physical origin of the idea, I discuss
its possible extension which may lead to new applications\footnote{The
history and recent developments in the subject were summarized briefly in
\cite{bialsum}.}.

A wounded source, by definition, emits a certain density of particles,
independently of the number of collisions it underwent inside the
nucleus. To explain the physical meaning of this concept, let us
recall that the idea originated from the observation that the process of
particle production is not instantaneous \cite{lanpom}. A simplified
version of the argument \cite{sto} can be presented as follows.

Consider a particle created in a high-energy collision. In the reference
frame where the longitudinal momentum of this particle vanishes, the
minimal time necessary for its creation is $t_{0}\geq 1/\m$ where
$\m=\sqrt {m^{2}+\p^{2}}$ is its energy. In the
''laboratory'' frame where the target nucleus is at rest, 
the particle in question acquires some longitudinal momentum, the time
is multiplied by  Lorentz factor, and we have
\begin{equation}
t\geq\gamma t_{0}=\frac{E}{\m^{2}}=\frac{\cosh y_{lab}}{\m},
\ll{t}
\end{equation}
where $E$ is the energy of the particle. Consequently, the uncertainty
of the distance from the collision point to that at which the particle
is created (i.e. the resolving power in the longitudinal distance) is
\begin{equation}
L=vt\geq\frac{\sinh y_{lab}}{m_{t}}. \label{l}%
\end{equation}
When the rapidity of the produced particle is large enough so that $L >
Z(b)$, where $Z(b)$ is the size of the nucleus at a given impact
parameter, the particle cannot resolve separate collisions and therefore
it is natural to suggest that its creation may be insensitive to the
number of collisions of the source. This is the origin of the concept of
a wounded source. One sees that it makes sense only for production of
particles with the laboratory rapidity exceeding that determined by the
condition $L > Z(b)$. 

Applications of this idea to "minimum bias" events dominated, as is
well-known, by production of pions at low transverse momentum
\cite{kitw} were met with a good deal of success \cite{bialsum}. A
particularly good description is obtained, within the quark-diquark
dominance picture \cite{bbz}, which may be considered as a
modification of  the dual parton model \cite{DPM}.

It is also well-known, however, that the model fails  for
production of heavy particles and/or particles having transverse
momentum exceeding $\sim 500$ MeV \cite{phen,npa}. At low energies this may have been
attributed to $m_\perp$ in the denominator of (\ref{l}), implying small
$L$ for larger $m_\perp$. But the data from RHIC proved without any
doubt that even when the condition (\ref{l}) is satisfied, production of
particles at high $m_\perp$ exceeds that predicted in \cite{bbz}
(for the review of data, see e.g. \cite{npa}).
 
In the present note I explore the possibility that the transverse mass
of a created particle, apart from defining  the minimal time needed for its
creation [c.f. (\ref{t})],  is also related to the transverse size $\d$ of
the source from which it is emitted. The idea is based on the observation that
 the quantum nature of the emission process suggests the  uncertainty relation 
\ba 
<\d ><\p>\simeq 1  \ll{dmt} 
\ea 

We shall investigate the consequences of this idea for 
the $A$-dependence of transverse momentum of  produced particles.

In the next section we remind briefly the quark-diquark model.
Generalization of the concept of wounded constituents implementing
(\ref{dmt}) is presented in Section 3. The formulae for the transverse mass spectra are derived in sections 4 and 5. The cross-sections of the wounded constituents are discussed in Section 6 together with some numerical exercises. Our conclusions are listed in the last section.  Application to the  Tsallis distribution is developed in the Appendix.

\section {Wounded nucleons, quarks and diquarks}

The beginning of the idea of wounded nucleons \cite{bbc} was purely
empirical. The first accelerator measurements of  multiplicities in
nucleon-nucleus collisions \cite{busza1} have shown that the average
multilicity follows the simple rule
\ba
n_{HA}= \frac12(\nu_A+1)n_{HH}  \ll{bb}
\ea
where 
\ba
\nu_A =\frac{A \sigma_{HH}}{\sigma_{HA}}
\ea
 is the average number of collisions of the projectile inside the 
nucleus. 

This result came as a surprize because "everybody" was expecting the
relation $n_{HA}= \nu_A n_{HH}$ which seemed much more natural, as it
suggests that each collision contributes approximately the same amount
to the observed particle multiplicity. The formula (\ref{bb}), on the
other hand, can be easily understood if one accepts  that 
each nucleon contributes the same amount, independently of the number of
collisions it suffered in the process.

Although the idea worked reasonably well for total multiplicities, the
understanding of the rapidity distributions came only 30 years later. To
make the long story short\footnote{A brief history of some of these
efforts can be found in \cite{bialsum}.}, let me just say that, as far
as I can judge, there were three essential steps: (i) the generalization
of the concept of wounded nucleons to that of wounded constituents
(originally: quarks \cite{bcf}, \cite{bc}-\cite{marco}, see also \cite{other}) which allowed to make the idea more
flexible,  (ii) abandoning the requirement of boost-invariance
\cite{bicz} and (iii) accepting that the contribution from a single wounded constituent is not restricted to one hemisphere \cite{bicz} (see also \cite{gg},\cite{ar}). Finally, a good description of ($p_\perp$ integrated) RHIC data at 200
GeV was obtained assuming that nucleon contains two independent sources
of particles: a constituent quarks and a constituent diquark \cite{bbz}.
Particle densities produced by quark and by diquark were assumed 
identical and could be determined from data. They turn out 
strongly asymmetric and thus obviously violating  boost-invariance. They are
not restricted to one hemisphere but extend throughout almost
full rapidity region,  in conformity with the results of  \cite{bicz}.

\section {Generalization: wounded constituents}

We have seen that the concept of wounded sources is well founded in the
basic theory and -at the same time- it is an useful tool in description
of data on particle production from various projectiles and targets. It
is also clear, however, that definite predictions can only be obtained
when the specific nature of these sources is precisely defined. Indeed,
the results from the wounded nucleon model are substantially different
from those of the wounded quark-diquark model and those differ, in turn,
from the wounded quark model. In short, the concept of wounded sources
must be supplemented by information on the nature of sources, about
their numbers and their  cross-sections. Only then the concept may
be effectively used to uncover the hidden relations between various
processes.

As already mentioned, the idea of wounded sources, as exploited till now \cite{bialsum}, shows one serious disadvantage. While it describes reasonably well the physics
at low $\p$, it fails badly at $\p$ exceeding 500 MeV and for heavy
particles \cite{phen,npa} where particle production increases with the size of the target faster than predicted by any wounded source model. At high energies (e.g. those of
RHIC) this failure cannot be attributed to the violation of the
coherence condition (\ref{l}). One sees therefore that some element of
the game is missing.

A hint can be obtained from comparison with data. It was recently
shown that the wounded nucleon model works very well for distributions in
the limit $\p\rr0$ \cite{ochs}. Furthermore, the data
integrated over $\p$ (i.e. dominated by $\p$ below $\simeq 300$ MeV) can be
described by the wounded quark-diquark model \cite{bbz}. Moreover, 
  it is well-established (e.g. from numerous experiments in deep inelastic lepton-hadron collisions) that the number, life times, energies 
and (transverse) sizes of the constituents in a hadron are by no means
fixed but are  distributed within a rather broad spectrum. Consequently,
in a collision of two hadrons various constituents may interact and get
"wounded". Each wounded constituent emits secondary particles and the final result is a sum of contributions from all of them. The number of wounded constituents of a given type depends on how many are present in the colliding hadrons as well as on
their corresponding cross-sections which in turn may depend on their
characteristics (e.g. colour charge, interaction strength and transverse
size).

In the present paper we show that this new picture radically changes the
predictions of the wounded constituent model for  particle production
at medium and large transverse momenta. Two effects are contributing to this 
result:

(i) Constituents of various energies and (transverse) sizes are expected to emit particles (mostly gluons)  
  with various distributions of transverse momenta and rapidities. Here we study the consequences of the simplest and natural choice, suggested by uncertainty principle:
\ba
\frac{dn(\p,\d;y)}{d^2\p dy} \equiv \rho (\p,\d;y)= I(\d;y) e^{-\p^2\d^2}  \ll{unc}
\ea
where $\d$ is the transverse size of the "source" (i.e. a constituent wounded in the collision). The intensity $I(\d,y)$ is independent of $\p$ but may depend on some other relevant variables as e.g.  mass of the emitted particle $ m$ and the  energy of the collision.

(ii) The cross-section $\s_{\d H}$ for the collision of a constituent  with a nucleon  is also expected to be sensitive to constituent's size. This in turn will influence the number of wounded constituents contributing to particle emission.

Using these ideas one can formulate the prediction of 
 the model of wounded constituents for the observed distribution in the collision of the nuclei $A$ and $B$ as 
\ba
\frac{dn_{AB}(\p,y)}{d^2\p dy}\equiv \rho_{AB}(\p,y)=\int dw_A(\d,b;B)\rho (\p,\d;y)+
\int dw_B(\d,b;A)\rho (\p,\d;-y)  \lb{mod}
\ea
where $dw_A(\d,b;B)$ is the number of constituents of size between $\d$ and $\d+d\d$ wounded in nucleus $A$ in the collision with the nucleus $B$ at the impact parameter $b$.

To make use of this prediction it is necessary to recall the old formula for the number of 
wounded constituents in a collision of two composite objects \cite{bbc,bcf}.  Consider a 
 collision of two nuclei $A$ 
 and $B$. For the number $dw_A$ of wounded
constituents of size between $\d$ and $\d+d\d$  in $A$ we have 
\ba
\sigma_{AB}(b)dw_A(b;\d;B)=\int dN_A(\d;s)\s_{\d B}(b-s)= AdN_H(\d)\int d^2s D_A(s) \sigma_{\d B}(b-s)\equiv \nn\equiv AdN_H(\d) \hat{\sigma}_{\d B}(b)\ll{wl}\hspace{6cm} \lb{dwlr}
\ea
and an analogous formula for $dw_B$. Here $ dN_H(\d)$ is the number of
constituents of size between $\d$ and $\d+d\d$ in the nucleon, $dN_A(b;s)=AD_A(s)dN_H(\d)$ is the number of constituents of size between $\d$ and $\d+d\d$ in the nucleus at the impact parameter $s$,  $D_A(s)$ is the (transverse) distribution of the nucleons in the nucleus $A$ normalized to unity,  $\sigma_{\d B}(b)$ is the cross-section of one constituent of size $\d$ on the nucleus $B$, and  
$\sigma_{AB}(b)$ is the total (inelastic) cross-section for the $A-B$
collisions\footnote{All cross-sections we discuss are understood as
inelastic, non-diffractive.}. 

\section {Nucleon-nucleon collisions}

The basic formula (\ref{mod}) contains a product of three, essentially unknown,  functions: the intensity  $I(\d;y)$,
the number $dN_H(\d)$ of the constituents inside a nucleon and the cross-section $\s_{\d B}(b)$.  In this section we show that this product can be derived from the existing data on transverse momentum distribution  in nucleon-nucleon collisions. 

For nucleon-nucleon collisions, Eq (\ref{wl}) gives for the number of wounded constituents  in one of them 
\ba
dw_H(\d) =dN_H(\d) \frac{\s_{\d H}}{\s_{HH}}.  \lb {dwh}
\ea
When this is inserted into (\ref{mod}) we obtain for the observed transverse momentum distribution in nucleon-nucleon collisions 
\ba
\rho_{HH}(\p;y)=\frac1{\s_{HH}}
 \int dN_H(\d)\s_{\d H}[ I(\d,y)+I(\d;-y)] e^{-\p^2\d^2}  \equiv \nn\equiv 
 \int [G(\d,y)+G(\d,-y)]e^{-\p^2\d^2} d\d =\int_0^\infty\frac{[G(\d,y)+G(\d,-y)}{2\d}e^{-\p^2\d^2} d\d^2     \lb{hh}
\ea 
with
\ba
G(\d,y)\equiv \s_{\d H}\frac{dN_H(\d)}{d\d} I(\d,y).
\ea
 Thus one sees  that the function $[G(\d,y)+G(\d,-y)]/\d$ can  be obtained from the measured transverse momentum distribution   by inverting the  Laplace transform (\ref{hh}).

In the present paper we take advantage of the observation that, for $p_\perp$ below 1-2 GeV,  the measured distributions of transverse momenta are well described by the exponential\footnote{For discussion of larger transverse momenta, see Appendix 1.}:
\ba
 \rho_{HH}(\p,y) = [n(y)+n(-y)]e^{-\beta \m}  \lb{pt}
\lb{ftau}
\ea
where $\beta$ is a constant and $n(y)$ describes the rapidity dependence. 

Combining (\ref{hh}) and (\ref{pt}) and using the identity  \cite{abra}
\ba
e^{-\beta \m}=\int_0^\infty du e^{-u\m^2}\frac{\beta }{2\sqrt{\pi u^3}}e^{-\beta^2/4u}=
\frac{\beta }{\sqrt{\pi}} \int \frac{ d\d}{\d^2}e^{-\beta^2/4\d^2} e^{-\m^2\d^2}
\ea 
we obtain \cite{fluc}
\ba
G(\d;y)d\d=n(y)\frac{\beta }{\d^2\sqrt{\pi}} e^{-m^2\d^2}
e^{-\beta^2/4\d^2}d\d \lb{dofdp}
\ea

At this point a remark is necessary. For the emission of gluons ($m$=0)  formula (\ref{dofdp}) for $G(\d;y)$ implies  that the nucleon contains very large constituents. Indeed, 
for $\d \rightarrow \infty$, $G$ falls only as $1/\d^2$, giving a really long tail, hardly acceptable.   Therefore an additional cut-off is necessary. We shall take it in the form
  $e^{-\mu^2\d^2}$ with $\mu\approx 1/2r_H$ where $r_H$ is the nucleon radius. This gives
 \ba
G(\d;y)d\d=n(y)\frac{\beta }{\d^2\sqrt{\pi}} e^{-\d^2/4r_H^2}
e^{-\beta^2/4\d^2}d\d \lb{dofdpg}
\ea
and thus the transverse momentum distributions  depend solely on  
\ba
\mm\equiv \sqrt{\p^2+\frac1{4r_H^2}}.
\ea

\section {Nuclear collisions}

For the nucleon-nucleus collision at the impact parameter $b$  one sees from  (\ref{wl}) that the number of wounded constituents in the nucleon is
\ba
\s_{HA}(b)dw_A(b,\d)= dN_H(\d) \s_{\d A} (b),
\ea
whereas the number of wounded constituents in the nucleus  is 
\ba
dw(b,\d)=\nu_A(b)dw_H(\d)  
\ea
where $\nu_A(b)$ is the number of nucleon-nucleon collisions at the impact parameter $b$: 
\ba
\nu_A(b)=A\frac{\s_{HH} D_A(b)}{\s_{HA}(b)}
\ea
and
$D_A(b)$ is the distribution of the nucleons in the nucleus $A$ (normalized to 1). 
Consequently, for the distribution of transverse momenta we obtain from (\ref{mod}):
\ba
\s_{HA}(b) \rho_{HA}(\p;b;y)=\nn= \int dN_H(\d)\s_{\d A}(b)I(\d;y)e^{-\p^2\d^2}d\d+
AD_A(b)\int dN_H(\d)\s_{\d H} I(\d;-y)e^{-\p^2\d^2}d\d =\nn=
\s_{HH} A\int d\d G(\d,y) \frac{\s_{\d A}(b)}{A\s_{\d H}}e^{-\p^2\d^2}+A
\s_{HH}D_A(b)n(-y)e^{-\beta\mm}.  \lb{haf}
\ea

Integration over impact parameter gives
\ba
 \rho_{HA}(\p;y)=\nu_A\left[\int d\d G(\d,y) \frac{\s_{\d A}}{A\s_{\d H}}e^{-\p^2\d^2}+
n(-y)e^{-\beta\mm}\right]     \lb{hafb}
\ea
where $\nu_A=A\s_{HH}/\s_{HA}$ is the number of collisions averaged over all impact parameters.

For nucleus-nucleus (A-B) collisions, Eqs. (\ref{mod}) and (\ref{wl}) give
\ba
\s_{AB}(b) \rho_{AB}(\p,b;y)=AB\s_{HH} \int \left\{ G(\d,y) \frac{\hat{\s}_{\d A}(b)}{A\s_{\d H}}+ G(\d,-y)\frac{\hat{\s}_{\d B}(b)}{B\s_{\d H}}\right\}
e^{-\p^2\d^2}d\d.  \lb{ab}
 \ea
 
When integrated over impact parameters, this formula gives
\ba
\rho_{AB}(\p;y)=\nu_{AB}\int \left\{ G(\d,y) \frac{\s_{\d A}}{A\s_{\d H}}+ G(\d,-y)\frac{\s_{\d B}}{B\s_{\d H}}\right\}
e^{-\p^2\d^2}d\d  \lb{abb}
 \ea
where $\nu_{AB}=AB\s_{HH}/\s_{AB} $ is the average number of nucleon-nucleon collisions.

Eqs. (\ref{haf})-(\ref{abb}) give the distribution  of transverse momentum of the observed particle. It is not difficult to see that the corresponding formulae for data integrated over some region of transverse momenta (from $\p^{(min)}$ to $\p^{(max)}$) are obtained by the simple substitution
\ba
G(\d,y)\;\rightarrow\; \pi G(\d,y)\frac{e^{-[\p^{(min)}]^2 \d^2}-e^{-[\p^{(max)}]^2 \d^2}}{\d^2}.
\ea 
This formula may be interesting for two reasons:

(i) Very often data are taken within a limited $\p$ range \cite{lhc};

(ii) In nucleus-nucleus collisions  substantial corrections to (\ref{ab}) and (\ref{abb}) are expected because the observed spectra are modified by the effects of the flow and  of the "jet quenching" \cite{npa}. These corrections are, however, much less important for data integrated over $d^2\p$ and thus such data may provide a more direct test of the model.

\section {Cross-sections and a numerical exercise}

One sees from the previous discussion that the only unknown in the problem is the cross-section $\s_{\d H}$ from which the ratio $\s_{\d A}/\s_{\d H}$ can be evaluated by standard methods. 

To have nuclear enhancement increasing  with increasing transverse mass (as observed experimentally for  $\m\leq 2$ GeV, the region of interest here),  the cross-section $\s_{\d A}/\s_{\d H}$ should be small at small $\d$. This can be naturally accommodated  if  the constituents we are dealing with are colour neutral and thus exhibiting the phenomenon  of colour transparency. Accepting this point of view (a possible interpretation  is discussed in the last section) we take, as a first choice,  the form used in \cite{gbw}, i.e.

 \begin{figure}[h]
\begin{center}
\includegraphics[scale=0.8]{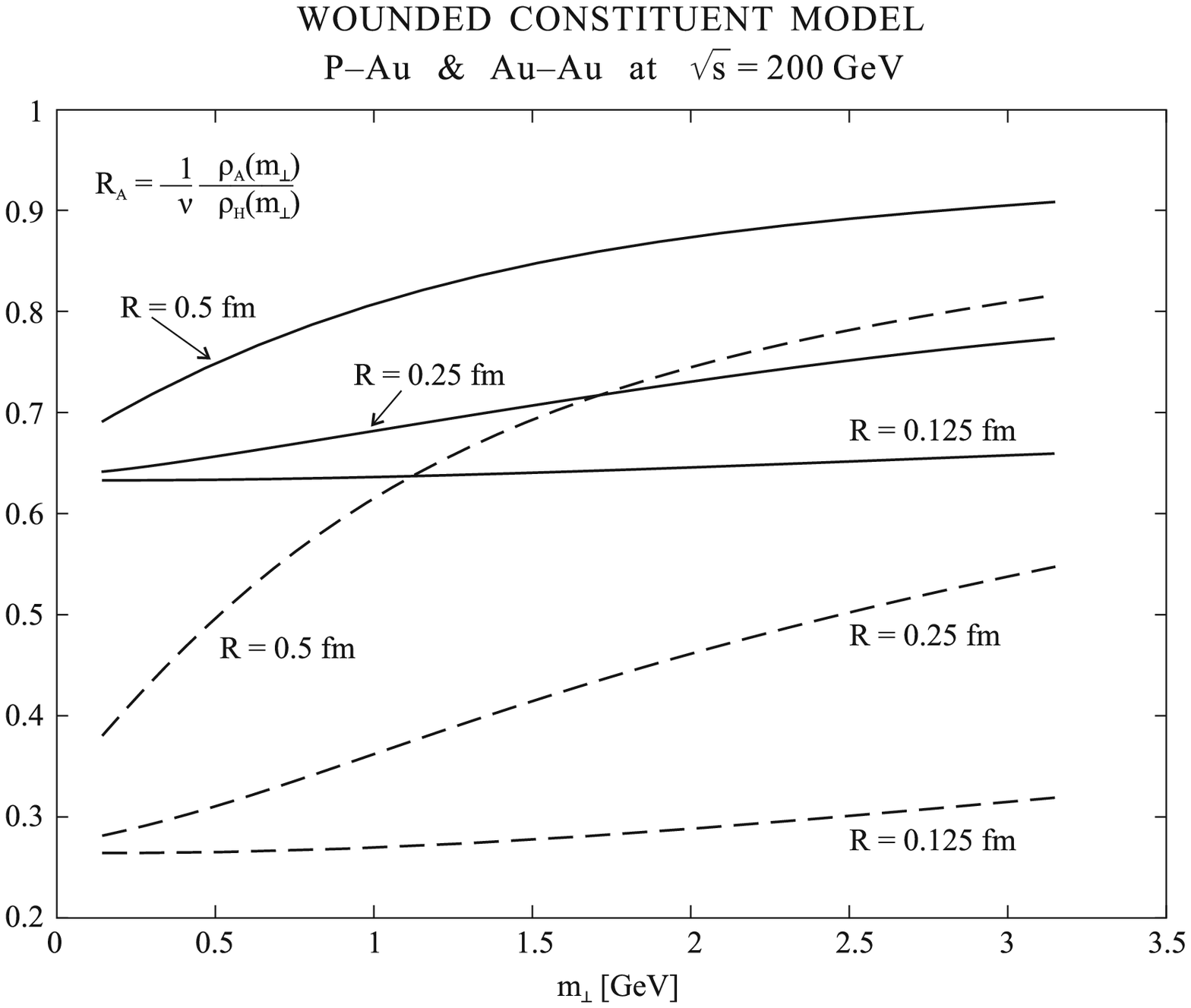}
\end{center}
\caption{The nuclear enhancement ratios $R_{HA}(\m)$ and $R_{AA}(\m)$ [(\ref{rha}),(\ref{raa})], plotted versus $\mm$ for various values of $R$, as indicated in the figure. Full lines: p-Au; Dashed lines: Au-Au.}%
\label{fig_rapt}%
\end{figure}

\ba
\s_{\d H}= \s_0\left[1-e^{-\d^2/R^2}\right]  \lb{gbsig}
\ea
where $\s_0$ and $R$ are parameters. This formula implies that for large $\d$ the cross-section saturates at the value $\s_0$. In this limit $\d\rightarrow \infty$ there is apparently only one constituent inside the nucleon and therefore one may expect that  the nuclear effects are identical to those of the wounded nucleon model.  Therefore, as a first approximation, we  take
\ba
\s_0\equiv \s_{HH}  \lb{ses}
\ea
Thus we are left with only one parameter, $R$, which determines how fast the constituent cross-section increases from zero to its limiting value\footnote{For the physical interpretaion of $R$, see \cite{gbw,nik}}. 
It should be  clear that, since $\s_{\d H}$ is always smaller than $\s_{HH}$, particle production in the present model is always larger than in the wounded nucleon model.

For illustration, and to obtain  a feeling how strong are the effects we are discussing and how sensitive  are they to the value of $R$, we evaluated the nuclear enhancement ratios from the formulae given in sections 4 and 5. We have taken $ r_H=0.7$  fm,  $\s_{HH} =30$ mb, $\s_{HAu}=1550$ mb, $\s_{AuAu}$ =2560 mb which seem appropriate at RHIC energy $\sqrt{s}=200$ GeV.
To avoid the problem of rapidity dependence we have considered $y=0$ where ${\bf n(y)}=n(-y)$ and thus $n(y)$ simply drops in the ratios.

\begin{figure}[h]
\begin{center}
\includegraphics[scale=0.8]{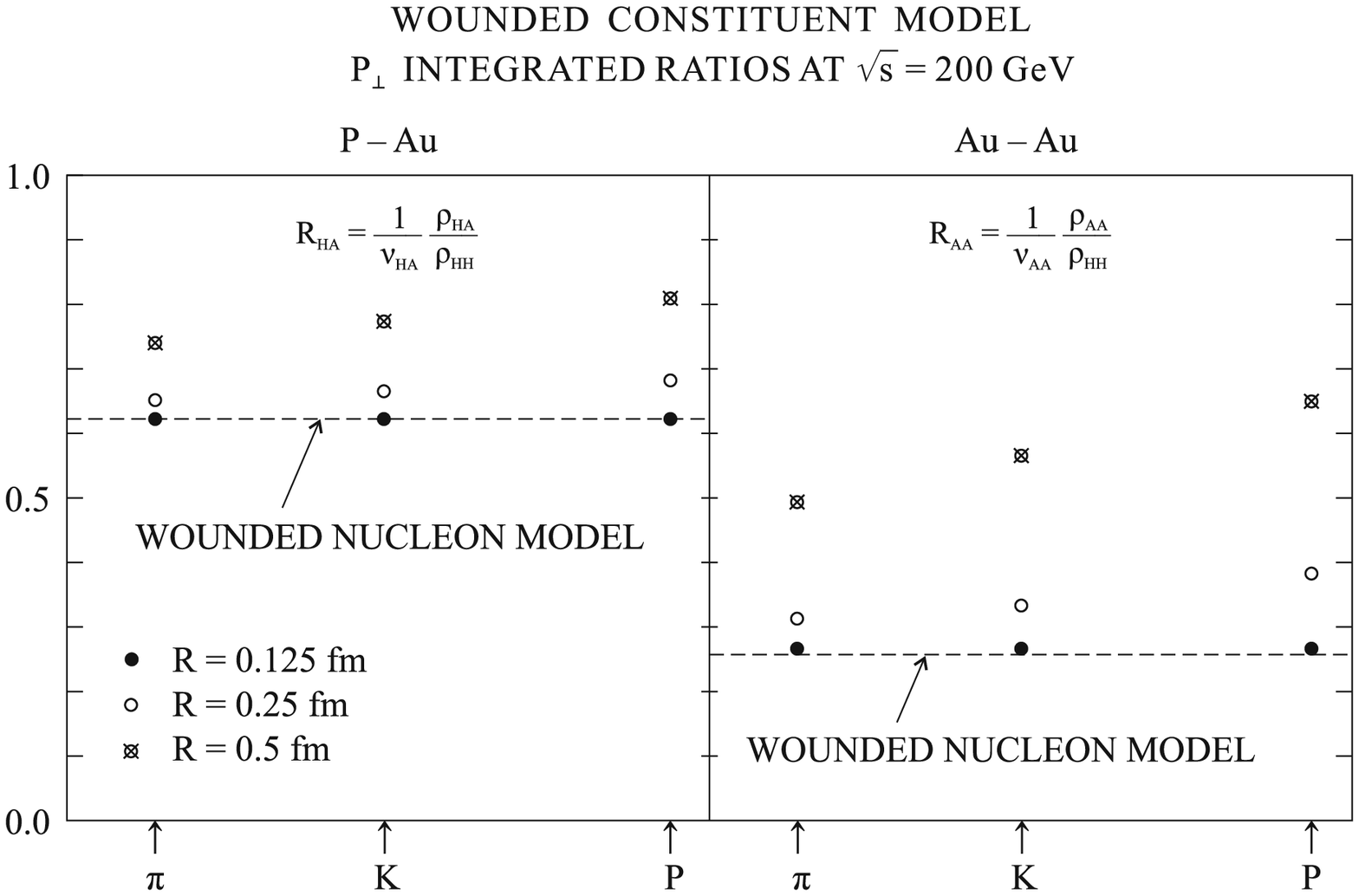}
\end{center}
\caption{The nuclear enhancement ratios $R_{HA}$ and $R_{AA}$ of yields for various particles, integrated over $d^2\p$, plotted for various values of $R$, as indicated in the figure.}%
\label{fig_ra}%
\end{figure}

In Fig. 1 the $\mm$ dependence of the ratios
\ba
R_{HA}(\p) =\frac{\s_{HA}}{A\s_{HH}} \frac{\rho_{HA}(\p)}{ \rho_{HH}(\p)}=\frac1{\nu_{HA}} \frac{\rho_{HA}(\p)}{ \rho_{HH}(\p)} \lb{rha}
\ea
 and
\ba
R_{AA}(\p) =\frac{\s_{AA}}{A^2\s_{HH}} \frac{\rho_{AA}(\p)}{ \rho_{HH}(\p)}=\frac1{\nu_{AA}} \frac{\rho_{AA}(\p)}{ \rho_{HH}(\p)} \lb{raa}
\ea 
 evaluated from Eqs. (\ref{hafb}) and (\ref{abb}) for p-Au  and Au-Au collisions
 is plotted versus $\mm$  for various values of  $R$, ranging from 0.125 to 0.5 fm.  
 One sees a strong dependence on $R$. At the smallest value, $R$=0.125 fm, the results are consistent with the wounded nucleon model. As $R$ increases, one observes a clear increase of $R_A$ with $\mm$ from the value close to that predicted by the wounded nucleon model at small $\mm$ towards the asymptotic value\footnote{Note that in this model  $R_A\leq 1$ for any $\p$. Thus  the Cronin effect \cite{cronin} cannot be described in this framework, unless the multiple scattering corrections are included \cite{wang}.} $R_A= 1$ at larger $\mm$.

 In Fig. 2 the  same ratios but for the spectra integrated over $\p$ [Eq. (\ref{hafb})] are plotted
 for pions, kaons, and nucleons\footnote{Here we assume that the $\m$ dependence of  the observed hadrons spectra (but  not their normalization, of course) are identical to the $\mm$ spectra of gluons.}.
 One sees that, as expected, the nuclear effects increase with increasing mass of the particle.

\section {Conclusions and outlook}

 The main purpose of this paper is to point out that the composite nature of hadrons, as revealed in numerous experiments of deep inelastic scattering, does have important consequences for the process of particle production at high energies. Following the old argument \cite{lanpom,sto}, based on the observation that particle production process in not instantaneous, the idea of the "wounded" constituents is formulated.  Its simplest consequences for nucleon-nucleus and nucleus-nucleus collisions were studied. It turns out that this approach has a good chance to improve significantly the description of
 the main features of these processes, particularly at transverse momenta exceeding 
 $\sim 200 $MeV.
 
In the present formulation  many potentially important effects were omitted either for the sake of clarity or because they require  more work. Their (most
 likely incomplete) list is given below.
 
 (i) Although, as argued in Section 2,   multiple scatterings inside the nucleus should not influence the rapidity distribution of  particles emitted from a source, they are expected to change the transverse momentum of the source and thus, in consequence, also the transverse momentum distribution of the emitted particles with respect to the direction of the projectile. It was shown \cite{wang} that  multiple scattering can be responsible for the Cronin effect \cite{cronin} observed in the proton-nucleus scattering. Thus it should  be included before serious comparison with data is done.
   
 (ii) In nucleus-nucleus collisions the collective phenomena  are changing the 
 observed spectra: the transverse flow modifies spectra at low transverse momenta while the effects of jet quenching influences the large transverse momentum tail. They have to be taken into account before the data are analyzed. These effects are much weaker (if present at all) in the nucleon-nucleus collisions which seem, therefore, a better place to 
 test the soundness of the ideas presented here.
 
 (iii) The corrections listed in (i) and (ii) influence mostly the transverse momentum distributions. If one integrates over the transverse momenta they are largely removed  and the result may be closer to reality. Such integration removes also, however, the  most spectacular predictions of the present approach.
 
 (iv) The emission from the "wounded" source is, most likely, dominated by gluons and thus the argument presented in this paper refers, at least formally, solely to gluon distributions. To discuss the actual particle spectra, the hadronization with all its complications has to be included. Hopefully, these effects may, at least partly,  cancel in the ratios of nuclear and nucleon particle yields.

Additional comments are in order.

(i) It may be interesting to speculate about the nature of the constituents we are considering. Accepting that a high-energy hadron is built  from quarks and gluons, it is natural to expect  the effect of {\it colour screening}, i.e. formation of  domains of various sizes in which the colour is locally compensated. These regions are of course fluctuating in size and in time but at very high energy they are frozen and can be treated as "constituents".  During the inelastic (i.e. colour-exchanging) collision  with the target     such a "constituent" becomes coloured (i.e. "wounded")  and starts to radiate gluons\footnote{It was pointed out already long time ago by G.Baym \cite{baym} that the very concept of a "wounded constituent" can be consistently defined only for colour-neutral objects.}. 

(ii) A progress could be obtained if one knew the proper normalization of the Eq. (\ref{unc}) i.e. the intensity of the emission from a Gaussian source. This is, however, a difficult problem which requires a separate study.

(iii) Our argument opens a way to more detailed investigation of the distribution and of the nature of the constituents forming the nucleon. For example, an attempt to derive (\ref{gbsig}) from the "elementary" cross-section of the two constituents could give some interesting clues about $dN_H(\d)$ [c.f. (\ref{dwlr}), (\ref{dwh})]. Also the relation between  the constituent cross-section  and the total nucleon-nucleon cross-section \cite{czm} can be used for this purpose.  All these problems, although interesting, fall far beyond the subject of the present investigation.

\bigskip

\textbf{Acknowledgements} This contribution is borrowing heavily from the
recent papers written together with Adam Bzdak. I would like to thank him
for very interesting discussions and for the long and fruitful
collaboration.  Thanks are also due to Krzysztof Fialkowski for helpful comments.
This investigation was supported in part by the grant N N202
125437 of the Polish Ministry of Science and Higher Education (2009-2012).

\section {Appendix : Tsallis distribution}

Extension of the analysis to larger transverse momenta demands more precise treatment of $\p$ distribution. The exponential (\ref{pt}) represents a reasonable approximation only 
at $\p$ below 1 GeV. At larger $\p$ the so-called Tsallis formula \cite{tsal}, which can be interpreted as a superposition of simple exponentials \cite{prl}, is more adequate \cite{pts}:
\ba
 \frac{dn^{HH}}{d^2p_\perp} = 
 \frac{n_0}{[1+\beta \m /k]^k}=n_0\frac{k^k}{\Gamma(k)}\int_0^\infty  t^{k-1} e^{-kt} 
 e^{-\beta\m t} \lb{ftau}
\ea
where $k$ and $\beta$ are  independent of $\m$ but may depend on other variables, 
e.g. the energy of the collision and rapidity. $n_0$ is the normalization factor, responsible for the total multiplicity.

Thus we have to find a function $G(\d)$ satisfying the condition\footnote{The parameters on the R.H.S. may depend on $m$, $y$ and the total energy.} 
\ba
 \int G(\d)  e^{-\p^2\d^2} d\d =n_0\left[1+\beta \m/k\right]^{-k}
\ea

 This can be done in two steps. First we convert $e^{-\m^2\d^2}$ into exponential, as shown in Section 4, and then use (\ref{ftau}) to obtain 
\ba 
G(\d) d\d=\frac{k^k}{\Gamma(k)}\frac{\beta }{\sqrt{\pi}} \frac {n_0}{\d^2}e^{-m^2\d^2}
d\d\int_0^\infty t^k dt  e^{-kt} e^{-\beta^2t^2/4\d^2} . \lb{dofd}
\ea

In the limit of $k\rightarrow\infty$ the Tsallis formula goes into exponential and we recover the formula for $D(\d)$ used in the main text.

\end{document}